\documentclass[12pt]{article}
\usepackage{amsmath}
\usepackage{amsbsy}
\usepackage{amssymb}
\usepackage{multirow}
\usepackage{esint}
\usepackage{graphicx}
\usepackage{float}
\usepackage[colorlinks= true, urlcolor= black, citecolor = black, linkcolor = blue]{hyperref}
\usepackage{caption}
\usepackage{subcaption}
\usepackage{setspace}
\usepackage[numbers]{natbib}
\usepackage{adjustbox}
\usepackage{graphicx,color}
\usepackage{lineno}

\begin{document}

\title{Robust Genomic Prediction and Heritability Estimation using Density Power Divergence}
\author{Upama Paul Chowdhury, Ronit Bhattacharjee,  Susmita Das and Abhik Ghosh 
\\ Indian Statistical Institute, Kolkata, India. }
\date{\today}
\maketitle

\begin{abstract}
This manuscript delves into the intersection of genomics and phenotypic prediction, focusing on the statistical innovation required to navigate the complexities introduced by noisy covariates and confounders. The primary emphasis is on the development of advanced robust statistical models tailored for genomic prediction from single nucleotide polymorphism data 
in plant and animal breeding and multi-field trials. 
{The manuscript highlights} the significance of incorporating all estimated effects of marker loci into the statistical framework and aiming to reduce the high dimensionality of data while preserving critical information. 
This paper introduces a new robust statistical framework for genomic prediction, 
employing one-stage and two-stage linear mixed model analyses along with utilizing the popular robust minimum density power divergence estimator (MDPDE) 
to estimate genetic effects on phenotypic traits. 
The study illustrates the superior performance of the proposed MDPDE-based genomic prediction and associated heritability estimation procedures 
over existing competitors through extensive empirical experiments on artificial datasets 
and application to a real-life maize breeding dataset.  
The results showcase the robustness and accuracy of the proposed MDPDE-based approaches, especially in the presence of data contamination,
emphasizing their potential applications in improving breeding programs and advancing genomic prediction of phenotyping traits. 
\end{abstract}


\newpage 
\section{Introduction}
\label{SEC:intro}

In the expansive landscape of statistical innovation, the intersection of genomics and phenotypic prediction constitutes a pivotal frontier. 
The rise of high-throughput genotyping technologies, such as the extensive utilization of single nucleotide polymorphism (SNP) data, 
is driving a significant shift in statistical methodologies towards revealing the genetic foundations of intricate traits. 
The present manuscript contributes to the growing need for sophisticated statistical models tailored to employ the richness of genomic data 
while navigating the complexities introduced by covariates, confounders, and different noises (contamination) in the data,
specifically with the objective of \textit{genomic prediction} (of phenotypic traits) using 
{data from randomly sampled individuals from a population}.

In the realm of statistical genomics, genome-wide association studies serve as 
an important scientific method for identifying genetic variations 
associated with specific traits, diseases, or conditions in both plant and animal breeding 
(Uffelmann et al., 2021 \citep{uffelmann2021genome}; Lourenço et al., 2020 \citep{lourencco2020robust}). 
This process relies on a plethora of statistical models and methods, each crafted to refine meaningful insights from vast genomic datasets. 
Unlike traditional marker-assisted recurrent selection, 
genomic prediction integrates all estimated effects of marker loci { to predict genomic breeding values of individuals}, regardless of their individual association with the phenotype.
Advanced statistical techniques enhance the accuracy of such genomic prediction, 
uncovering the acute genetic effects often overlooked by conventional methods in handling extensive genetic markers. 
Many of these techniques aim to reduce high-dimensionality while preserving critical information (Giraud, 2021 \citep{giraud2021introduction}; 
Wainwright, 2019 \citep{wainwright2019high}; Buhlmann et al., 2014 \citep{buhlmann2014high}). 
Some models, however, do take into account the high-dimensional nature of the marker matrix and attempt to fit the entire genome 
using methods such as penalized-likelihood or Bayesian shrinkage estimation. 
However, several issues have been identified when employing these models for effect estimation including computational intensity and modeling complexity. 
Shen et al.~(2013) \citep{shen2013novel} suggested using heteroscedastic marker variances in ridge regression as an alternative to Bayesian genome-wide prediction models.
Hofheinz and Frisch (2014) \citep{hofheinz2014heteroscedastic} demonstrated the superior effectiveness of the heteroscedastic effect model 
by employing two methods within a one-stage linear mixed model analysis, 
encompassing all random effects (genetic effects, block effects, replicate effects, etc.) following classical literature. 
Alternatively, to avoid the computational complexity of this one-stage method, many authors proposed a second two-stage mixed model framework, 
where the random components are partitioned into two distinct stages; firstly genetic effects are considered 
as fixed effects while block and replicate effects are considered as random effects, 
and subsequently, in the second stage, genetic effects are regarded as random effects 
(Piepho et al., 2012 \citep{piepho2012efficient}; Estaghvirou et al., 2013 \citep{estaghvirou2014influence}). 
These types of mixed models have been commonly applied in plant  breeding techniques and multi-environment trials in previous literature 
(Westhues et al., 2021 \citep{westhues2021prediction}; Tanaka, 2020 \citep{tanaka2020simple}; Resende, 2016 \citep{resende2016software}; 
Bernal-Vasquez et al., 2016 \citep{bernal2016outlier}; Estaghvirou et al., 2015 \citep{estaghvirou2015genetic}; Falke et al., 2014 \citep{falke2014genome}; 
Shen et al., 2013 \citep{shen2013novel}; Piepho, 2009 \citep{piepho2009ridge}; Möhring and Piepho, 2009 \citep{mohring2009comparison}; 
Barbosa et al., 2005\citep{barbosa2005selection}). 
\\

In the present study, we develop a new robust statistical procedure for genomic prediction under possible \textit{data contamination}, 
based on complex  noisy data, effectively capturing complex interactions among genetic markers, environmental factors, and covariates. 
For these, we employ two model-based approaches, namely the one-stage and two-stage linear mixed model analyses from the literature of genomic prediction,
but propose to use the popular robust minimum density power divergence (DPD) estimation instead of classical maximum likelihood based estimation procedure.
The minimum DPD  estimator (MDPDE), after originally introduced by Basu et al. (1998) \citep{basu1998robust}, 
has subsequently gained widespread popularity due to its strong robustness properties with just a little loss in (asymptotic) efficiency, 
computational simplicity, and its capacity to offer a direct interpretation akin to the maximum likelihood estimator (MLE);
see, e.g., Basu et al. (2011) \citep{basu2011statistical}, Ghosh and Basu (2013) \citep{ghosh2013robust} and the references therein. 
Saraceno et al. (2020) \citep{saraceno2020robust} have recently studied the MDPDE for linear mixed effect models, 
which we employ here for developing robust genomic prediction under one and two stage mixed model approaches.	 
As a result, our proposed methods withstand influential factors, including the presence of outliers in various directions
and similar issues stemming from model misspecification, providing improved performances over the latest existing procedures of genomic prediction, 
as illustrated through extensive Monte-Carlo simulations with both uncontaminated and contaminated data. 
Furthermore, we evaluate applicability and performances of these proposed methods with a real dataset of a maize (\textit{Zea mays}) breeding study,
taken from Hofheinz and Frisch (2014) \citep{hofheinz2014heteroscedastic}.

Additionally, in the present study, we propose a robust procedure for estimating heritability, as a by-product of genomic prediction, 
that leads to more accurate result than traditional heritability estimators especially in the presence of outliers (or other noises/contamination) in the data.
Heritability is inherently entangled to bioethics. 
{The} advent of SNP-heritability warrants a renewed assessment, 
quantifying the genetic contribution to phenotypic variation within a population (Zhu and Zhou, 2020 \cite{zhu2020statistical}). 
We assess the effectiveness of the proposed robust heritability estimator also through Monte Carlo simulation studies 
and by applying them to an actual maize dataset mentioned above.

\section{Methodology}

\subsection{Mathematical Model for Genomic Prediction}
\label{SEC:model}

Suppose that our data consist of $N$ individual phenotype observations in $r$ replicate and $p$ biallelic single-nucleotide polymorphism (SNP) markers 
where genotype effects are distributed in $B$ blocks and $L$ confounder or controllers in a randomized complete block design.
The block often represents different crop types or soil/environmental conditions in plant breeding and multi-environment trials,
and confounding/controlling variables are any additional covariates 
(e.g., replication indicator, surrounding temperature/humidity, soil properties, etc., when they are not used to construct the blocks)
that we wish to include in our analyses as {random} effects 
from subject-matter knowledge on the actual experimental setting.
Mathematically, let $ \boldsymbol{y}_k$ denotes the vector of $n_k$ phenotype data for the $k$-th replicate for each $k=1, \ldots, r$, 
$ \mathbf{X}_g $ denotes the matrix of $p$ SNP markers along the genome for $N$ individuals 
($\mathbf{X}_g$ has $p$ columns for the SNP observations, usually coded as 0, 1, $-$1 for homo zygote aa, the hetero-zygote Aa and the other homo-zygote AA, respectively) 
$\mathbf{X}_b$ denotes the design matrix for random block effect and $\mathbf{Z}$ denotes the $N\times L$ matrix of confounders.
We denote the $k$-th partition of rows of $\boldsymbol{Z}$, $ \mathbf{X}_g $ and $\mathbf{X}_b$ associated with each $ \boldsymbol{y}_k$ 
by $\boldsymbol{Z}_k$, $\boldsymbol{X}_{gk}$ and $\boldsymbol{X}_{bk}$, respectively, for all $k=1, \ldots, r$. 
Following preceding works, e.g., Tanaka (2020) \citep{tanaka2020simple}, Lourenço et al.~(2020) \citep{lourencco2020robust}, 
Estaghvirou et al.~(2014) \citep{estaghvirou2014influence}, and Piepho et al.~(2012) \citep{piepho2012stage}, 
let us assume the a linear mixed model (LMM) for SNP data in the $k$-th replicate as given by
\begin{equation}\label{EQ:LMM0}
\boldsymbol{y}_k=\mathbf{Z}_k \boldsymbol{\gamma}+ \mathbf{X}_{gk} \mathbf{u}_g+\mathbf{X}_{bk} \mathbf{u}_b +\boldsymbol{\epsilon}_k, 
~~~~k=1, \ldots, r, 
\end{equation}
where  the $p$-vector of random SNP effect $\mathbf{u}_g \sim N(\mathbf{0}, \boldsymbol{\Sigma}_g)$,
the random block effects $\mathbf{u}_b \sim N(\mathbf{0},\sigma_{b}^{2}, \mathbb{I}_B)$, $\boldsymbol{\gamma}$ denotes the $l$-vector of the fixed effects
and the random error $\boldsymbol{\epsilon}_k \sim N(\boldsymbol{0},\sigma_{e}^2 \mathbb{I}_{n_k})$ for $k=1, \ldots, r$. 
Note that $N = \sum_{k=1}^{r} n_k$ which is the total number of observation and individual replication is assumed to be independent of each other;
the data are partitioned by replicates to utilize this independence across replications in the mathematical formulation of our robust estimation procedures. 

Putting  $ \boldsymbol{y}=(\boldsymbol{y}^{'}_1, \boldsymbol{y}^{'}_2,....,\boldsymbol{y}^{'}_r)^{'}$ to denote the vector of all phenotype data,
we can rewrite the LMM in (\ref{EQ:LMM0}) as 
\begin{equation}\label{EQ:LMM1}
\boldsymbol{y}
=
\mathbf{Z} \boldsymbol{\gamma}
+
\begin{bmatrix}
	\mathbf{X}_g & \mathbf{X}_b
\end{bmatrix}
\begin{bmatrix}
	\mathbf{u}_g \\ 
	\mathbf{u}_b \\
\end{bmatrix}
+
\boldsymbol{\epsilon}
= \mathbf{Z} \mathbf{\gamma} + \mathbf{X} \mathbf{u} + \boldsymbol{\epsilon},
\end{equation}
where $\boldsymbol{\epsilon}=(\boldsymbol{\epsilon}^{'}_1, \ldots, \boldsymbol{\epsilon}^{'}_r)^{'}$, $ \mathbf{X}=  \begin{bmatrix}
\mathbf{X}_g & \mathbf{X}_b 
\end{bmatrix} $ and
$ \mathbf{u}= \begin{bmatrix}
\mathbf{u}_g \\ 
\mathbf{u}_b 
\end{bmatrix}$.
Based on the observed data, our aim is to first estimate  the fixed effect $\boldsymbol{\gamma}$ and the random effect $\mathbf{u}$
and then use them for genomic prediction and heritability estimation.

\subsection{Estimating Marker Effects and Breeding Values}
\label{SEC:Est_MLE}

For ordinary Ridge regression the fixed effect $\boldsymbol{\gamma}$ and the random effect $\mathbf{u}$ can be estimated jointly via Henderson's mixed model equation (Henderson 1953) given by
\begin{equation}\label{EQ:MME}
\begin{bmatrix}
	{\textbf{Z}}^{T} \mathbf{Z} & \textbf{Z}^{T}\mathbf{X}\\ \textbf{X}^{T}\mathbf{Z} & \textbf{X}^{T}\mathbf{X} + \mathbf{\Lambda} 
\end{bmatrix}
\begin{bmatrix}
	\boldsymbol{\gamma} \\ \textbf{u}
\end{bmatrix}
=
\begin{bmatrix}
	\textbf{Z}^{T}\boldsymbol{y} \\ \textbf{X}^{T}\boldsymbol{y}
\end{bmatrix},
\end{equation}
where $\boldsymbol{\Lambda} = Diag(\lambda_1,\lambda_2,.....,\lambda_p)$, with $\lambda_j$ being the shrinkage parameter of $j$-th SNP for $j=1, \ldots, p$, 
which is indeed  taken as a function of  $\sigma_{e}^{2}$  and $\boldsymbol{\Sigma}_g$.

For estimating the marker-effects, the traditional approach is to minimize the likelihood based loss function associated with the model (\ref{EQ:LMM1}) 
which can be simplified to have the form 
$$
L(\mathbf{u},\boldsymbol{\gamma})=(\boldsymbol{y}-\mathbf{Xu}-\mathbf{Z}\gamma)^{T}(\mathbf{y-Xu-Z}\gamma)+\mathbf{u}^{T}\Lambda \mathbf{u}.
$$
This loss function can be minimized analytically to get an estimator of the random-effect $\mathbf{u}$  
(including both SNP-marker effects $\mathbf{u}_b$ and block effects $\mathbf{u}_b$) as given by
\begin{equation}\label{EQ:uhat}
\widehat{\mathbf{u}}= \begin{bmatrix}
	\widehat{\mathbf{u}}_g \\ 
	\widehat{\mathbf{u}}_b 
\end{bmatrix} = (\mathbf{X}^{T}\mathbf{M_{Z}X}+\boldsymbol{\Lambda})^{-1}\mathbf{X}^{T}\mathbf{M_{Z}y},
\end{equation}
where $\mathbf{M_Z}=\mathbb{I}_N - \mathbf{Z}(\mathbf{Z}^{T}\mathbf{Z})^{-1}\mathbf{Z}^{T}$ 
is the projection matrix removing the effect of confounding or controller variables,
$\boldsymbol{\Lambda}$ is a diagonal matrix of shrinkage parameters $\lambda_j$, $j=1, \ldots, p$.
These shrinkage parameters $\lambda_j$, being a function of the variance parameters, can be estimated from different types of variance estimators and 
subsequently we can get the estimated $\widehat{\mathbf{u}}$ from (\ref{EQ:uhat}) using estimated values of $\lambda_j$s in $\boldsymbol{\Lambda}$. 
Then, for genomic prediction, the estimate of breeding value will be given by \citep{shen2013novel}
\begin{equation}\label{EQ:est_breeding}
\widehat{\mathbf{g}}=\mathbf{X}_g \cdot \widehat{\mathbf{u}_g},
\end{equation}
and the estimated phenotype effects can be obtained as 
\begin{equation}\label{EQ:est_pheno}
\widehat{\textbf{y}}=\mathbf{Z} \cdot \widehat{\boldsymbol{\gamma}} + \mathbf{X} \cdot \widehat{\mathbf{u}}. 
\end{equation}

We would like to emphasize that there exist several approaches for obtaining the estimates of random effects 
and error variances leading to different estimates of the shrinkage parameters  in the literature of genomic prediction. 
For ordinary ridge regression, we assume the simplest form $\boldsymbol{\Sigma}_g = \sigma_{g}^{2} \mathbb{I}_p$ 
and then term $\boldsymbol{\Lambda}$ can be replaced by $\lambda \mathbb{I}$ with the constant shrinkage parameter being computed as 
$\widehat{\lambda}= \frac{\widehat{\sigma}_e^{2}}{\widehat{\sigma}_g^{2}}$.
However, while using the likelihood based estimation approaches, it has been shown that the heteroscedastic values of $\lambda_j$ works significantly better  
where we assume $\boldsymbol{\Sigma}_g = Diag\left(\sigma_{g1}^{2}, \ldots, \sigma_{gp}^{2}\right)$ and estimate different shrinkage parameters, corresponding to different SNPs, as 
$\widehat{\lambda}_j=\frac{\widehat{\sigma}_e^{2}}{\widehat{\sigma}_{gj}^{2}}$ for every $j=1, \ldots, p$.
The most common one is to use the restricted maximum likelihood estimator under the assumed LMM which is often referred to as the ridge regression (RR) estimator. 
Hofheinz and Frisch (2014) \cite{hofheinz2014heteroscedastic} proposed two methods, to be referred to as RMLV and RMLA, 
that are shown to have significantly improved performances over the classical RR estimator for genomic prediction;
they both estimate the random effect $\widehat{\mathbf{u}}$ using (\ref{EQ:uhat}) with $\lambda_j$ replaced by its estimators $\widehat{\lambda}_j$ for all $j$ 
under the model assumption $\boldsymbol{\Sigma}_g = Diag\left(\sigma_{g1}^{2}, \ldots, \sigma_{gp}^{2}\right)$, 
but these  $\widehat{\lambda}_j$ are obtained in two ways as described below:
\begin{itemize}
\item RMLA: First, a moment estimator of marker-specific variance component is obtained from random single-factor analysis of variance (ANOVA) for $j$-th marker as given by
\begin{equation}
	\widehat{\sigma}_{gj}^{2*}=\frac{MQM_j - MQE_j}{\frac{1}{2}\cdot(N-\sum_{i}{\frac{n_{i}^{2}}{N})}},
\end{equation}
where $ MQM_{j}$  denotes the mean square due to $j$-th SNP marker, $MQE_{j}$ denotes the error for the $j$-th SNP marker, 
and $n_{i}$ is the numbers of individuals in the $i$-th SNP-marker factors for $i=1, 2, 3$ corresponding to three SNP-marker factors 0, 1, $-$1, respectively.
Then, the shrinkage parameters are estimated from the resulting variance estimates as given by 
\begin{equation}
	\widehat{\lambda}_{j} = \frac{\sigma_{e}^{2}}{\sigma_{u}^{2}} \cdot \frac{\sum_{j}{\widehat{\sigma}_{gj}^{2*}}}{\widehat{\sigma}_{gj}^{2*}}, 
	~~~~~~~j=1, \ldots, p.
\end{equation}
Here, $\sigma_{u}^{2}$ total genetic variance estimated from the model (\ref{EQ:LMM1})

\item RMLV: This method is a modification of the restricted maximum likelihood procedure, producing heteroscedastic SNP-variance through appropriate expectation-maximization algorithm.  
In this approach, we first obtain the estimates of the error variance and the marker variance for the $j$-th SNP marker (for each $j=1, \ldots, p$) as
$$
\widehat{\sigma}_{e}^{2}=(\textbf{y}^{'}\textbf{y}-\widehat{\boldsymbol{\gamma}^{'}}\textbf{Z}^{'}\boldsymbol{y}-\widehat{\textbf{u}^{'}}X^{'}\textbf{y})/(N-1), 
~~~\mbox{	and   } ~~~
\widehat{\sigma}_{gj} ^{2}=(\widehat{\textbf{u}_j}^{'}\widehat{\mathbf{u}_j} - \sigma_e^{2} \cdot tr(\mathbf{C}_{jj}))/j,
$$
where $tr(\mathbf{C}_{jj})$ is the trace of inverse of the left most (coefficient) matrix in (\ref{EQ:MME}). 
Then the  shrinkage parameters are estimated as  $ \widehat{\lambda}_j=\frac{\widehat{\sigma_e}^{2}}{\widehat{\sigma}_{gj}^{2}}$ for each $j=1, \ldots, p$. 
\end{itemize}
These two approaches, RMLV and RMLA, are considered as robust by the authors in \cite{hofheinz2014heteroscedastic} since the RMLV is robustified by taking different marker-variance 
and the RMLA is robustified with both different marker-variance and shrinkage parameters for each marker effects. 
However, these estimators are seen to be highly unstable against data contamination or noises in the observed datasets, which is the primary focus of the present work. 
Our proposed estimation approaches based on the minimum DPD estimation, that yield robust results in the presence of data contamination, are described in the following subsection.

\subsection{The proposed Robust Estimation Approaches}
\label{SEC:est_MDPDE}

\subsubsection{The Minimum DPD Estimation for General set-up}
\label{SEC:MDPDE}

The minimum DPD estimator (MDPDE) was initially developed by Basu et al.~(1998)\cite{basu1998robust} for independent and identically distributed data,
say  $\textbf{y}_1, \textbf{y}_2, \ldots,\textbf{y}_n$ from a true distribution $G$, to be modelled by a parametric family of distributions indexed by $\boldsymbol{\theta}\in\Theta$. 
This approach relies on the minimization of a divergence measure, namely the DPD, 
between the true density $g$ and the model density $f_{\boldsymbol{\theta}}$ to obtain a robust and highly efficient estimate of the unknown parameter vector $\boldsymbol{\theta}$. 
This particular divergence is defined in terms of a tuning parameter $\alpha\geq 0$  as follows 
$$
d_\alpha(g,f_{\boldsymbol{\theta}}) =
\begin{cases}
\int f_{\boldsymbol{\theta}}^{\alpha+1} - (1+(1/\alpha)) \int f_{\boldsymbol{\theta}}^\alpha g + \frac{1}{\alpha} \int g^{1+\alpha}, & \alpha > 0 \\
\int g \ln \big(\frac{g}{f_{\boldsymbol{\theta}}}\big), & \alpha=0.
\end{cases}
$$
In practice, based on the observed data, we replace the unknown density $g$ suitably by the help of the empirical distribution function estimate 
which leads to the following simplified objective function for obtaining the MDPDE as (see, e.g.,  Basu et al., 1998 \cite{basu1998robust})
$$
\int f_{\boldsymbol{\theta}}^{(1+\alpha)} - \left(1+\frac{1}{\alpha}\right) \frac{1}{n} \sum_{i=1}^n f_{\boldsymbol{\theta}}^\alpha (\textbf{y}_i).
$$
This leads to robust parameter estimates for $\alpha>0$ with the tuning parameter $\alpha$ controlling the trade-off between 
the (asymptotic) efficiency under pure data and the robustness under contaminated data. 
It can be seen that the MDPDE at $\alpha=0$ coincides (in a limiting sense) to the most efficient by highly non-robust maximum likelihood estimator (MLE)
as the DPD measure at $\alpha=0$ is indeed the Kullback-Leibler divergence.
As $\alpha>0$ increases, the corresponding MDPDE achieves greater robustness with a little loss in its pure data (asymptotic) efficiency;
see Basu et al. (2011) \cite{basu2011statistical}  for more details about the MDPDE along with its applications under standard model set-ups.

The theory of the MDPDE has been extended for the independent but non-homogeneous data by Ghosh and Basu (2013) \citep{ghosh2013robust} and subsequently found many applications including regression and mixed models. 
Whenever our observed data $\textbf{y}_1,\textbf{y}_2, \ldots,\textbf{y}_n $ are independent but non-homogeneous with each $\textbf{y}_i \sim g_i$ for all $i=1, \ldots, n$ 
and $g_i$ is being modeled by the parametric density $f_i(\cdot, \boldsymbol{\theta})$, the MDPDE of the common unknown parameter $\boldsymbol{\theta}$ 
can be obtained by minimizing the average divergence between the data points and the respective model densities; 
this becomes equivalent to the minimization of a simpler objective function given by (Ghosh and Basu, 2013 \citep{ghosh2013robust})
\begin{equation}\label{EQ:MDPDE_objFunc1}
H_n(\boldsymbol{\theta}) = \frac{1}{n} \sum_{i=1}^n \bigg[\int f_i(\textbf{y},\boldsymbol{\theta})^{(1+\alpha)}dy - \left(1+\frac{1}{\alpha}\right)f_i(\textbf{y}_i,\boldsymbol{\theta})^{\alpha}\bigg] =   \frac{1}{n} \sum_{i=1}^n H_i(\textbf{y}_i,\boldsymbol{\theta})
\end{equation}
Once again the case $\alpha=0$ coincides with the MLE and we get improved robust estimators at larger values of $\alpha>0$ with only a little loss in efficiency;
see, e.g., Ghosh and Basu (2013, 2015, 2016) \citep{ghosh2013robust,ghosh2015robust,ghosh2016robust} and Ghosh (2019) \cite{ghosh2019robust}  
among many others, for further details and applications. In particular, Saraceno et al.~(2020) \citep{saraceno2020robust} used this general theory of the MDPDE 
to develop robust estimators under the general set-up of linear mixed-models 
which we would be using in our context to get robust estimators of marker effects and breeding values and eventually a robust genomic prediction procedure.

\subsubsection{The MDPDE based One-stage Approach for Genomic Prediction}
\label{SEC:MDPDE_one}

For robust genomic prediction in the one-stage approach, we consider the mathematical model described in (\ref{EQ:LMM0}) with $\boldsymbol{\Sigma}_g = \sigma_{g}^{2} \mathbb{I}_p$
and propose to estimate the associated parameters robustly using the MDPDE instead of the likelihood based approach described in Section \ref{SEC:Est_MLE}.
In the context of genomic predictions, from the assumed model (\ref{EQ:LMM0}), 
we get that the $n_k$-vector $\boldsymbol{y}_k$ containing the target phenotype measurements from the $k$-th replicate 
has the distribution  
$\boldsymbol{y}_k \sim N(\mathbf{Z}_k\boldsymbol{\gamma},\mathbf{V}_k)$ independently for each $k=1, \ldots, r$, where
$$
\boldsymbol{V}_k=\sigma_e^2 \mathbb{I}_{n_k} + (\boldsymbol{X}_g \boldsymbol{X}_g^{T} \sigma_{g}^2 +\boldsymbol{X}_b \boldsymbol{X}_b^T \sigma_{b}^2).
$$  
This set-up clearly belongs to the independent non-homogeneous set-up described in the preceding subsection and 
hence the robust MDPDE of the model parameters  $ \boldsymbol{\theta} = (\boldsymbol{\gamma}', \sigma_g^2, \sigma_{b}^2, \sigma_e^2)'$ 
can be obtained by minimizing the objective function given in (\ref{EQ:MDPDE_objFunc1}) which, for the present case, further simplifies to
\begin{equation}
	H_n(\boldsymbol{\theta})= \frac{1}{r} \sum_{k=1}^r  \frac{1}{(2\pi)^{\alpha} |\mathbf{V}_k|^{\alpha/2}} \left[\frac{1}{(\alpha +1)}-\left(1+\frac{1}{\alpha}\right)
	e^{\frac{\alpha}{2}(\boldsymbol{y}_k-\mathbf{Z}_k\boldsymbol{\gamma})^T \textbf{V}_k^{-1} (\boldsymbol{y}_k-\mathbf{Z}_k\boldsymbol{\gamma})}\right].
\end{equation}

In our implementation, we have minimized the above objective function numerically in \texttt{R} to obtain the MDPDE of model parameter in our numerical exploration presented in this paper.
One can alternatively compute these MDPDEs by solving the associated estimating equations obtained as 
$ \frac{\partial {H_n(\boldsymbol{\theta})}}{\partial \boldsymbol{\gamma}} = \boldsymbol{0}$ 
and $\frac{\partial {H_n(\boldsymbol{\theta})}}{\partial \boldsymbol{\delta}}=0 $ with $\boldsymbol{\delta} = ( \sigma_{g}^2, \sigma_{b}^2,\sigma_e^2)$.
We refer to Saraceno et al.~(2020) \citep{saraceno2020robust} for more details including the asymptotic and robustness properties of the resulting parameter estimates.

Once we have obtained the robust MDPDE of the variance parameters as $\widehat{\sigma}_g^{2}$ and $\widehat{\sigma}_e^{2}$, 
we can use them to get an estimate of $\boldsymbol{\Lambda}$ as in the case of ordinary ridge regression, 
by assuming $\boldsymbol{\Lambda}=\lambda \mathbb{I}$ and estimating $\lambda$ by $\widehat{\lambda}= \frac{\widehat{\sigma}_e^{2}}{\widehat{\sigma}_g^{2}}$.
Then the random effects can be robustly estimated from (\ref{EQ:uhat}) with $\boldsymbol{\Lambda}=\widehat{\lambda} \mathbb{I}$ 
and subsequently  be used for robust estimation of the breeding values and prediction of phenotype effects following (\ref{EQ:est_breeding}) and (\ref{EQ:est_pheno}), respectively.

\subsubsection{The MDPDE based Two-stage Approach for Genomic Prediction}
\label{SEC:MDPDE_two}

For robust genomic prediction in the two-stage approach, we start with the mathematical model described in (\ref{EQ:LMM1}) 
with $\boldsymbol{\Sigma}_g = \sigma_{g}^{2} \mathbb{I}_p$
and split it in two sequential models following Estaghvirou et al.~(2013) \citep{estaghvirou2015genetic} and Piepho et al.~(2012) \citep{piepho2012stage}. 
In the first stage, we consider a simpler model for the observed phenotype response as given by 
\begin{equation}\label{EQ:LMM2.1}
	\boldsymbol{y}=\boldsymbol{\eta}\boldsymbol{\mu} + \mathbf{X}_b \mathbf{u}_b+\boldsymbol{\epsilon}.
\end{equation}
As described in Piepho et al. (2012) \citep{piepho2012stage}, the matrix $\boldsymbol{\eta} $ and the vector $\boldsymbol{\mu}$
can be constructed in such way that 
$ \boldsymbol{\eta} \cdot \boldsymbol{Z}_1 = \boldsymbol{Z} $ and $ \boldsymbol{\eta} \cdot \boldsymbol{X}_1 = \boldsymbol{X}_g $
and then $\boldsymbol{\mu} = \boldsymbol{Z}_1 \boldsymbol{\gamma} + \boldsymbol{X}_1 \boldsymbol{u}_g $
for some suitably defined matrices $\boldsymbol{Z}_1$ and $\boldsymbol{X}_1$.
Note that, $\boldsymbol{\eta}$ may be thought of as the fixed-effect design matrix with associated parameter vector $\boldsymbol{\mu}$
in this simpler first-stage model (\ref{EQ:LMM2.1}). 
Here, both the marker effects ($\mathbf{u}_g$) and the genotypic mean ($\boldsymbol{\gamma}$)
are considered as fixed parameters within the adjusted mean $\boldsymbol{\mu}$ 
and only the block effects ($\mathbf{u}_b$) are treated as random effects in the model (\ref{EQ:LMM2.1}).
Following Lourenco et al. (2020) \citep{lourencco2020robust}, we would use the robust M-estimators with Huber's weight function, as proposed and studied in Koller (2013), in this first stage modeling to estimate $\boldsymbol{\mu}$ 
and associated variance components $\sigma_b^2$ and $\sigma_e^2$.
These estimates are computed using the function \texttt{rlmer} from the \texttt{R}-package 
\texttt{robustlmm} (Koller, 2016 \cite{koller2016robustlmm}) in our numerical experiments.

Once we get the estimated adjusted mean as $ \widehat{\boldsymbol{\mu}}$, it is then used as response in the second stage model given by 
\begin{equation}\label{EQ:LMM2.2}
	\widehat{\boldsymbol{\mu}}=\mathbf{Z}_1 \boldsymbol{\gamma} + \mathbf{X}_1 \mathbf{u}_g + \widetilde{\boldsymbol{\epsilon}}, 
\end{equation}
where $\widetilde{\boldsymbol{\epsilon}}$  denotes the vector of (new) random errors having variance-covariance matrix same 
as that of the adjusted mean $\boldsymbol{\mu}$ from the first stage analysis. 
Assuming normality of these new error terms, we can now estimate $\boldsymbol{\gamma}$, $\sigma_g^2$ 
and the new error variance parameter (say $\widetilde{\sigma}_e^2$) by a suitable method. 
We propose to use the robust MDPDE for this purpose which can be obtained by solving the associated objective function as described previously. 
Once we get the robust estimates of all the parameters, we can obtain the robust estimates of the breeding values and prediction of phenotype effects as before 
following Eqns. (\ref{EQ:est_breeding}) and (\ref{EQ:est_pheno}), respectively.

\subsection{Measuring Heritability}
\label{SEC:Est_Heritability}

The extent to which genetic differences contribute to the observed variability in a particular trait within a population is measured by SNP-based heritability.
{Mathematically, it can be measured by the ratio of genetic variance $(\sigma_{g}^2)$ to the total phenotypic variance ($\frac{\widehat{\sigma}_e^{2}}{r}$)}, so that	it lies between $0$ to $1$ 
An estimate of the heritability of a trait is specific for populations as well as environment and changes with circumstances. 
Once we get the robust MDPDE of the variance parameters as $\widehat{\sigma}_g^{2}$ and $\widehat{\sigma}_e^{2}$, 
we can estimate the heritability following 
{as 
	\begin{equation}
		\widehat{H}_p = \frac{\widehat{\sigma}_g^2}{\widehat{\sigma}_{g}^2 + \frac{\widehat{\sigma}_e^{2}}{r}}.
\end{equation}}
Note that, a high value 0.5 of $\widehat{H}_p$ implies that on average half of the differences among phenotypes depend on genetic effect whereas a value close to 0 indicates that almost all of the variability in a trait among population are not under the genetic control and due to some other factors such as environment.
We will empirically illustrate that use of the proposed MDPDE based variance estimators 
leads to a much stable and accurate heritability estimator than the existing measures.

\section{Experiments}

We will illustrate the superior performance of the proposed MDPDE based one and two-stage genomic prediction and heritability estimation over its recent competitors  
through extensive empirical experiments involving artificial datasets from different known model setups as well as a real-life breeding dataset.
These simulation models and datasets are described in the following subsections, 
along with the performance metrics and the competing methods considered in all our experiments.

\subsection{Experimental Setup: Artificial Datasets}

We consider the simulation model setups resembling real-life situations similar to the one used in Lourenco et al. (2020) \citep{lourencco2020robust}. 
The underlying true data-generating model is taken as 
\begin{equation}
	y_{ijk} = \phi + g_i + b_{jk} + e_{ijk},
\end{equation}
where $y_{ijk}$ is the phenotypic trait value for $i$-th genotype in the $j$-th block within $k$-th {replicate},
$\phi$ is the overall mean, $g_i=\sum_{l=1}^{p}{z_{il}u_{g_l}}$  be the $i$-th genotypic value with $z_{il}$ being the SNP-marker values
and $u_{g_l}$ being the random effects generated from $N(0, \sigma_g^2)$ independently for all $j=1, \ldots, p$, 
$b_{jk}$ is random block effect of $j$-th block of $k$-th replicate, 
computed based on random effects associated with $B$ blocks generated independently from $N(0, \sigma_b^2)$ along with the block design matrix,
and $e_{ijk}$ is random error generated from $N(0, \sigma_e^2)$ independently for all $i, j, k$.   

\textcolor{black}{
In order to perform the simulations following a realistic genetic architecture, commonly used in plant breeding scenarios,
we have used the \texttt{R}-package \texttt{AlphaSimR} \cite{gaynor2021alphasimr}. 
First, we have simulated the haplotype sequence of the founder population using the \textit{MaCS} (Markovian Coalescent Simulator) program \cite{chen2009fast} that is implemented within the \texttt{AlphaSimR} package; 
the parameter setup in simulating the founder population are taken from \cite{lourencco2020robust,muleta2019optimizing}.
In particular, we have taken $N = 715$ individuals where each individuals have 10 chromosomes with 11640 segregating sites per chromosome.
Subsequently, we have set our global parameters to model a single trait, utilizing the \texttt{R}-function \texttt{SimParam}, 
assuming 1000 QTLs per chromosome and SNP markers to be evenly distributed across the 10 chromosomes. 
The heritability value is set at 0.7 resembling the work done in Muleta et al. ~\cite{muleta2019optimizing}. 
Then, we have used the function \texttt{newPop} to simulate our initial population based on the founder haplotypes. 
The  SNP-marker matrix (with elements \( z_{il} \)) is generated using the function \texttt{pullSnpGeno} 
available within the package \texttt{AlphaSimR}; 
the SNP marker values are kept within \( \{0,1\} \), where $0$ is kept as homo-zygotes aa and AA, and $1$ is kept for hetero-zygote Aa. 
Finally, the function \texttt{bv}  from the same package \texttt{AlphaSimR} is used to generate our respective breeding values.
}

\textcolor{black}{
Each artificial dataset is generated following this model set-up with $k=2$ replicates, $B=70$ blocks and $p=11646$ SNP markers,
where 5 blocks contain 13 observations each and remaining 65 blocks contain 10 observations each, 
resulting in the total of $N = 715$ observations. 
For simulating  the block effects, we have utilized the \texttt{design.rcbd} and \texttt{design.alpha} functions 
from the \texttt{R}-package \texttt{Agricolae} \cite{de2019package}.
The parameter values are taken as considered in the study by Lourenco et al.~\cite{lourencco2020robust};
in particular, we take \(\phi=0.05\), \(\sigma^2_r=0\), \(\sigma^2_b=6.27\),\(\sigma^2_e=53.8715\) and \(\sigma^2_g=0.005892\). 
}

Additionally, to illustrate the claimed robustness properties, we also generate artificial contaminated datasets from each simulated pure datasets (as described above) based on the following two contamination schemes: 
\begin{itemize}
	\item Random contamination: For each dataset, 5\% phenotype observations are randomly chosen and replaced by its original value plus 5 times of the standard deviation of the residual error.
	\item Block contamination: For each dataset, 5 blocks are chosen randomly and the phenotype observations under those blocks (for all replications) are replaced by its original value plus 8 times of the standard deviation of the residual error.
\end{itemize}

\subsection{Competitive Methods}

For one-stage approach, we consider the methods described as RMLA and RMLV in Section \ref{SEC:Est_MLE} 
since they are shown to outperform other ridge regression based method by Hofheinz and Frisch (2014) \citep{hofheinz2014heteroscedastic}.
However, since our primary focus is on robust inference against data contamination, we also compare our MDPDE based one-stage proposal 
with robust variants of RMLA and RMLV methods, which we would refer to as Rob-RMLA and Rob-RMLV; they are computed as in RMLA and RMLV, respectivley,
but using the Huber's M-estimation approach (Koller, 2013 \citep{koller2013robust}) 
to estimate the parameters of the underlying linear mixed models instead of the likelihood based approach. 
These robust M-estimators are considered as classical benchmark against our proposed MDPDE based approach and are computed using the function \texttt{rlmer} from the \texttt{R}-package \texttt{robustlmm} (Koller, 2016 \citep{koller2016robustlmm}). 

For two-stage approach, we consider two existing robust competitors based on the work of Lourenço et al.~(2020) \citep{lourencco2020robust} 
which we refer to as Rob1 and Rob2. 
Both these methods robustify only the first stage estimation by using the M-estimation approach as described in Section \ref{SEC:MDPDE_two};
but Rob1 uses the classical likelihood based estimation in the second stage modeling while Rob2 uses the robust M-estimation approach also in the second stage modelling. 
Recall that our proposal involves using the novel MDPDE based approach in the second stage modeling, while keeping the the first stage the same as Rob1 and Rob2.

\subsection{Performance Metrics}

In order to compare our proposed robust procedures with other existing benchmark methodologies of genomic predictions, 
we illustrate how close the predicted values of the phenotype traits ($\widehat{\boldsymbol{y}}$) 
are in comparison to their original values ($\boldsymbol{y}$). 
So, for each artificial datasets, we compute the following two measures to illustrate the accuracy of  genomic predictions:
\begin{itemize}
	\item Pearson correlation coefficient $\widehat{\rho} = \frac{Cov(\boldsymbol{y} ,\widehat{\boldsymbol{y}})}{\sqrt{Var(\boldsymbol{y}) Var(\widehat{\boldsymbol{y}})}}$.
	Note that, the higher the value of $\widehat{\rho}$ higher is the predictive accuracy. 
	This measure has been traditionally used in the related literature on genomic prediction (e.g., Lourenco et al., 2020 \citep{lourencco2020robust}). 
	
	\item Median absolute deviation (MAD): a robust accuracy measure obtained as the median of the absolute difference 
	between the predicted and the original values of the target phenotype traits. 
	A low value of MAD close to zero indicates good performance.
	We use this measure to get a robust indication of deviations in the predicted trait values.    
\end{itemize}
We compute one value of $\widehat{\rho}$ and MAD for each artificial dataset in our simulations using cross-validation, 
with a random 70-30 split into training and test data,
and report their average over 100 simulation replications in Tables \ref{tab:1}--\ref{tab:2}. 
These results represent the out-of-sample prediction accuracy of different methods, obtained based on the test data. 

Further, for each artificial dataset we get one final estimated value of the heritability measures $\widehat{H}_p$.
So, its performances under different existing and proposed approaches are examined by computing its Mean-square deviations (MSD) against its true value (known for each simulation setups)
over 100 simulation replications (Table 3).

\subsection{Real Data Analysis}

We consider a real life breeding dataset generated by the International Maize and Wheat Improvement Center (CIMMYT) 
that contains 300 tropical maize lines in 10 blocks. 
Genotyping was conducted using 1135 SNP markers and the target phenotypic trait was grain yield assessed 
under severe drought stress and well-watered conditions. These two conditions are treated as two replicates in our analyses. 

Within the 300 tropical maize lines, 16 lines have missing values in some SNP markers. 
To avoid computational challenges due to missing values, these datapoints are excluded from our analysis presented in this paper. 
Thus all methods of genomic prediction, both existing and proposed, and associated heritability estimation are applied to 
the remaining 284 datapoints to predict their grain yields from 1135 SNP markers suitably adjusted from the block and the replicate effects. 

In order to examine the performance of our proposed method for this real dataset,
we have again computed the correlation coefficient $\widehat{\rho}$ and the MAD between the original and the predicted values of the target phenotypic trait, 
which is grain yield in our datset. The resulting values of $\widehat{\rho}$ and MAD, 
{with 70-30 cross-validation as before}, and the estimated heritability measure $\widehat{H}_p$, 
are reported in Table \ref{tab:data} for different existing and proposed approaches applied on the test data;
corresponding results for the training data are given in the Appendix.

\section{Results and Discussions }

\subsection{Performance of the Proposed One-stage Method}

Prediction accuracy measures, namely $\widehat{\rho}$ and MAD, obtained for different one-stage procedures on the test data 
are presented in Table \ref{tab:1} for artificially generated pure data as well as two contaminated scenarios; 
similar results obtained for training data are given in Table \ref{tab:1A} in the Appendix. 
Examining these results, we observe that the  predictive accuracy of the proposed MDPDE based approach improves 
($\widehat{\rho}$ increases significantly and MAD mostly decreases) as $\alpha$ increases 
for both pure and contaminated situations.

\begin{table}[!h]
	\color{black}
	\centering
	\caption{Prediction accuracy measures obtained from different methods under the one-stage approach with artificial datasets 
		(test data).}
		\begin{tabular}{|l|c|c|c|c|c|c|}
			\hline
			\multirow{2}{*}{Methods} &  \multicolumn{2}{c|}{Pure Data}  &  \multicolumn{2}{c|}{Random Contamination} & \multicolumn{2}{c|}{Block Contamination} \\
			& $\widehat{\rho}$ & MAD & $\widehat{\rho} $ & MAD & $ \widehat{\rho} $ & MAD\\
			\hline
			RMLA 			& 0.532 & 23.84 & 0.270 & 18.38 & 0.322 & 11.12 \\ 
			RMLV 			& 0.587 & 16.77 & 0.387 & 16.69 & 0.470 & 18.38 \\ \hline
			Rob-RMLA 		& 0.540 & 25.10 & 0.343 & 17.50 & 0.520 & 13.29 \\ 
			Rob-RMLV 		& 0.712 & 7.23 & 0.430 & 8.18 & 0.550 & 6.48 \\ \hline
			\multicolumn{7}{|l|}{Proposed MDPDE based approach} \\ 
			$\alpha = 0.1$ 	& 0.667 & 6.02 & 0.620 & 7.34 & 0.667 & 6.34 \\ 
			$\alpha = 0.3$ 	& 0.745 & 6.30 & 0.650 & 7.37 & 0.645 & 6.37 \\ 
			$\alpha = 0.5$ 	& 0.737 & 6.67 & 0.537 & 6.67 & 0.537 & 5.67 \\ 
			$\alpha = 0.7$ 	& 0.710 & 5.37 & 0.630 & 6.37 & 0.210 & 5.37 \\ 
			$\alpha = 1$ 	& 0.823 & 4.37 & 0.670 & 5.30 & 0.440 & 5.13 \\ \hline
		\end{tabular}
		\label{tab:1}
\end{table}

In the context of pure data, the proposed MDPDE based prediction method  clearly outperforms other existing methods, 
namely both RMLA and RMLV and their robust counterparts, particularly in terms of MAD.
In terms of $\widehat{\rho}$, the performances of RMLV and Rob-RMLV are similar and better than RMLA and Rob-RMLA, respectively, 
and also competitive with the MDPDE based methods with larger values of $\alpha$. 
But, they are significantly worse compared to our proposed MDPDE based approach 
which can achieve a significantly small MAD value between 4.37--6.67 at any $\alpha>0$.

Under contamination, the performances of classical method of RMLA and RMLV deteriorate drastically 
while those of their robust counterparts (Rob-RMLA and Rob-RMLV) remain somewhat stable. 
The proposed MDPDE based procedure with suitable $\alpha>0$ clearly emerges as the superior choice in terms of 
both predictive accuracy measures 
with improved performances as $\alpha$ increases.

Comparing all set-ups and accuracy measures, the MDPDE based procedure remains most stable, 
illustrating their significantly improved robustness properties over the existing procedures.

\begin{table}[!b]
	\centering
	\color{black}
	\caption{Prediction accuracy measures obtained from different methods under the two-stage approach with artificial datasets 
		(test data).}
		\begin{tabular}{|l|c|c|c|c|c|c|}
			\hline
			\multirow{2}{*}{Methods} &  \multicolumn{2}{c|}{Pure Data}  &  \multicolumn{2}{c|}{Random Contamination} & \multicolumn{2}{c|}{Block Contamination} \\
			& $\widehat{\rho}$ & MAD & $\widehat{\rho} $ & MAD & $ \widehat{\rho} $ & MAD\\
			\hline
			Rob1 			& 0.754 & 4.10 & 0.740 & 5.5 & 0.710 & 5.29 \\ 
			Rob2 			& 0.793 & 4.23 & 0.730 & 4.18 & 0.740 & 4.48 \\ \hline
			\multicolumn{7}{|l|}{Proposed MDPDE based approach} \\ 
			$\alpha = 0.1$ & 0.820 & 4.02 & 0.760 & 4.14 & 0.747 & 4.34 \\ 
			$\alpha = 0.3$ & 0.845 & 4.20 & 0.750 & 3.67 & 0.742 & 4.37 \\ 
			$\alpha = 0.5$ & 0.837 & 3.67 & 0.737 & 3.39 & 0.733 & 4.67 \\ 
			$\alpha = 0.7$ & 0.810 & 2.37 & 0.820 & 3.08 & 0.758 & 3.37 \\ 
			$\alpha = 1$   & 0.860 & 2.37 & 0.832 & 2.93 & 0.800 & 3.34 \\ \hline
		\end{tabular}
		\label{tab:2}
\end{table}

\subsection{Performance of the Proposed Two-stage Method}

Table \ref{tab:2} presents the results of the predictive accuracy measures $\widehat{\rho}$ and the MAD 
for both pure and contaminated data with the two-stage modeling approach under test data; 
the same obtained for training data are given in Table \ref{tab:2A} in the Appendix. 
Based on the tabulated values, it is evident that the MDPDE method with $\alpha\geq 0.5$ provides 
the best estimates among all the second stage methods 
and the effect of $\alpha$ is almost negligible on their performances although $\alpha=1$ gives marginally best results. 
Further, the results obtained by two-stage procedures are also significantly better than the one-stage results as intuitively expected;
our proposed MDPDE based two-stage procedure can achieve a high correlation value of 0.86, 
indicating strong correlation between the predicted and the original phenotype traits (response). 
Our method also leads to predictions with smaller deviations (MAD) from their original values 
as compared with the existing Rob1 and Rob2 procedures. 
Further, the results from Table \ref{tab:2} (and Table \ref{tab:2A}) also indicate that, 
when MDPDE is used in the second stage of the two-stage modeling, 
there is a significant reduction in MAD values and an increase in estimated correlation coefficients as compared with the existing methods
for both block and response contaminated datasets. 
These observations suggest that the proposed MDPDE-based method exhibits even greater robustness 
compared to the only existing robust procedures.

\subsection{One-stage vs.~Two-stage Modeling}
\label{SEC:1vs2}

In this paper, we have examined two distinct types of model analysis, namely one-stage and two-stage modeling approaches. 
Based on the empirical analyses, it becomes evident that the two-stage approach yields superior predictive outcomes compared to the one-stage approach
as the incorporation of different random effects in two different stages enhances the robustness of marker-effect estimates.

\begin{figure}[!b]
	\begin{subfigure}[b]{0.5\textwidth}
		\includegraphics[width=\textwidth]{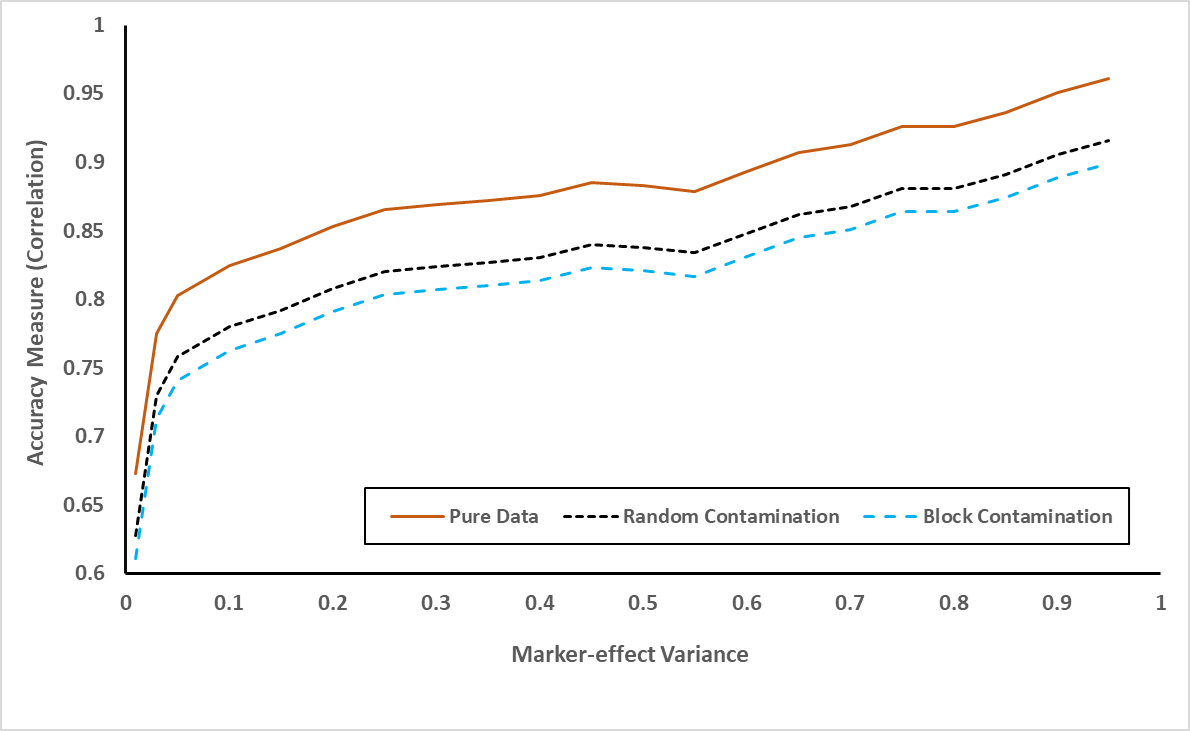}
		\caption{$\sigma_e^2 = 20.5$}
	\end{subfigure}
	\hfill
	\begin{subfigure}[b]{0.5\textwidth}
		\includegraphics[width=\textwidth]{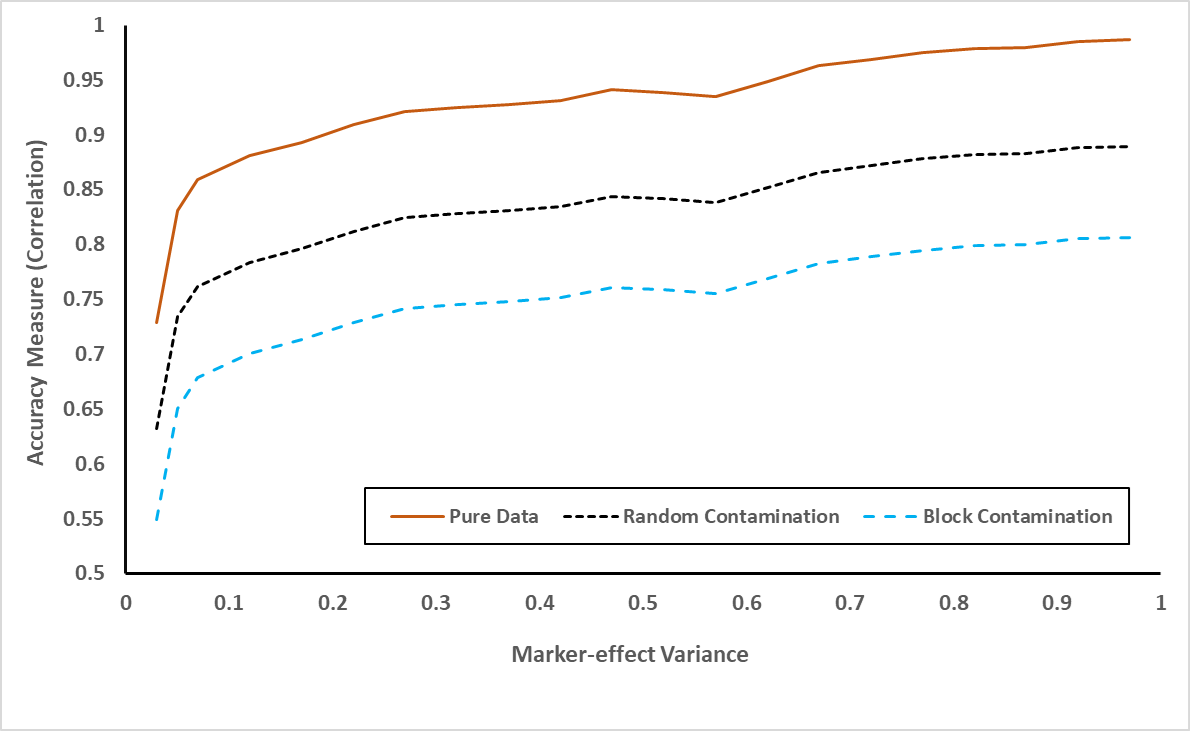}
		\caption{$\sigma_e^2 = 50.5$}
	\end{subfigure}
	\caption{Plots of the accuracy measure $\widehat{\rho}$ obtained by the MDPDE-based two-stage genomic prediction (with $\alpha=1$)
		over different SNP marker-effect variance ($\sigma_g^2$) in the second experimental setup for two different $\sigma_e^2$ (test data results).}
	\label{Fig:1vs2}
\end{figure}

However, it is to be noted that we have employed two different simulation setups for illustrating the accuracy of one-stage and two-stage modeling approaches. 
This is because, as we learned from extensive empirical explorations, performances of the proposed MDPDE based robust two-stage approach 
relies heavily on both error variance ($\sigma_e^2$) and (random) SNP-marker-effect variance ({$\sigma_g^2$}). 
To illustrate this dependence, in Figure \ref{Fig:1vs2}, 
we have plotted the correlation measure ($\widehat{\rho}$) between the estimated and the original phenotype traits
under the second experimental setup with different values of $\sigma_e^2$ and $\sigma_g^2$ for both pure and contaminated cases.
It can be clearly seen from these figures (and many similar numerical explorations not presented here for brevity)
that the accuracy of the proposed method increases as $\sigma_g^2$ increases and $\sigma_e^2$ decreases for both pure and contaminated datasets, 
which is equivalent to say that the genotypic signal-to-noise ratio of the underlying model increases. 
Therefore, for datasets with a high genotypic effect variance compared to the noise variance,  
genomic prediction using the proposed MDPDE based two-stage modeling approach will be more accurate and robust. 
In other cases with lower SNP effect variance or higher noise variances, the use of robust one-stage procedure is expected to produce better results.

\begin{table}[!b]
	\centering
			\label{tab:3}
	\color{black}{
		\caption{The average value and the MSD of heritability estimators obtained from different one-stage and two-stage procedures 
			applied on pure and contaminated artificial (test) data.}
		\resizebox{.95\columnwidth}{!}{
\begin{tabular}{|l|rr|rr|rr|}
	\hline
	{Methods} & \multicolumn{2}{|c|}{{Pure Data}} & \multicolumn{2}{|c|}{{Random Contamination}} & \multicolumn{2}{|c|}{{Block Contamination}} \\ 
	& {Average} & {MSD} & {Average} & {MSD} & {Average} & {MSD} \\ \hline
	\multicolumn{7}{|c|}{\textbf{One-stage procedures}} \\ \hline  
RMLA & 0.4295 & $7.3170 \times 10^{-4}$ & 0.4211 & $7.7785 \times 10^{-4}$ & 0.4028 & $8.8327 \times 10^{-4}$ \\ 
RMLV & 0.5673 & $1.7609 \times 10^{-4}$ & 0.5543 & $2.1228 \times 10^{-4}$ & 0.5210 & $3.2041 \times 10^{-4}$ \\ \hline
Rob-RMLA & 0.6473 & $2.7772 \times 10^{-5}$ & 0.5984 & $1.0322 \times 10^{-4}$ & 0.5672 & $1.7635 \times 10^{-4}$ \\ 
Rob-RMLV & 0.6732 & $7.1824 \times 10^{-6}$ & 0.6467 & $2.8408 \times 10^{-5}$ & 0.6115 & $7.8322 \times 10^{-5}$ \\ \hline
	\multicolumn{7}{|l|}{Proposed MDPDE based approach} \\
$\alpha = 0.1$ & 0.7653 & $4.264 \times 10^{-5}$ & 0.7276 & $7.6176 \times 10^{-6}$ & 0.7112 & $1.2544 \times 10^{-6}$ \\ 
$\alpha = 0.3$ & 0.7575 & $3.3062 \times 10^{-5}$ & 0.7348 & $1.211 \times 10^{-5}$ & 0.7154 & $2.3716 \times 10^{-6}$ \\ 
$\alpha = 0.5$ & 0.7701 & $4.914 \times 10^{-5}$ & 0.7365 & $1.3322 \times 10^{-5}$ & 0.7261 & $6.8121 \times 10^{-6}$ \\ 
$\alpha = 0.7$ & 0.8021 & $1.0424 \times 10^{-4}$ & 0.7985 & $9.7022  \times 10^{-5}$ & 0.7038 & $1.444 \times 10^{-7}$ \\ 
$\alpha = 1$ & 0.7843 & $7.1064  \times 10^{-5}$ & 0.7597 & $3.564  \times 10^{-5}$ & 0.7387 & $1.4976 \times 10^{-5}$ \\ \hline
	\multicolumn{7}{|c|}{\textbf{Two-stage procedures}} \\ \hline  
Rob1 & 0.7227 & $5.1529 \times 10^{-6}$ & 0.7041 & $1.681 \times 10^{-5}$ & 0.5768 & $1.5178 \times 10^{-4}$ \\ 
Rob2 & 0.8022 & $1.0444 \times 10^{-4}$ & 0.7836 & $6.9889 \times 10^{-5}$ & 0.6363 & $4.0576 \times 10^{-5}$ \\ \hline
	\multicolumn{7}{|l|}{Proposed MDPDE based approach} \\
$\alpha = 0.1$ & 0.8190 & $1.4161 \times 10^{-4}$ & 0.8030 & $1.0609 \times 10^{-4}$ & 0.7260 & $6.76 \times 10^{-6}$ \\ 
$\alpha = 0.3$ & 0.8230 & $1.5129 \times 10^{-4}$ & 0.8050 & $1.1025 \times 10^{-4}$ & 0.7332 & $1.1022 \times 10^{-5}$ \\ 
$\alpha = 0.5$ & 0.8488 & $2.2141 \times 10^{-4}$ & 0.8102 & $1.2144 \times 10^{-4}$ & 0.7229 & $5.2441 \times 10^{-6}$ \\ 
$\alpha = 0.7$ & 0.8670 & $2.2788 \times 10^{-4}$ & 0.8254 & $1.5725 \times 10^{-4}$ & 0.7481 & $2.3136 \times 10^{-5}$ \\ 
$\alpha = 1$   & 0.8900 & $3.61 \times 10^{-4}$ & 0.8543 & $1.543 \times 10^{-3}$ & 0.7570 & $3.249 \times 10^{-5}$ \\ \hline
\end{tabular}
}		}
\end{table}

\subsection{Accuracy and Robustness of the Heritability Estimator}

In Table 3, we present \textcolor{black}{the average values and} 
the mean square deviation (MSD) of the heritability estimate ($\widehat{H}_p$), 
obtained by different existing and proposed methods in one-stage and two-stage approaches 
based on artificial pure and contaminated datasets (test data). 
Corresponding results obtained under the training data are reported in Table 7 in the Appendix. 
In this respect, the proposed MDPDE based methods performs quite competitively to the existing robust methods,
for both one-stage and two-stage approaches, under pure data as well as under randomly contaminated data.
However, the MDPDE based methods show somewhat significant improvement in estimating the heritability over the existing methods
under the block contamination scenario.

\subsection{Comparative Performances for the Maize Dataset}

We have presented the correlation measure $\widehat{\rho}$ between the predicted and the true responses and their MAD 
obtained by different one-stage and two-stage methods in Table \ref{tab:data} along with the estimated values of the heritability measure ($\widehat{H}_p$).
In case of this real data, we can see that the proposed MDPDE based one-stage modeling approach yields the best fit in terms of both correlation and MAD;
the prediction performance further improves as $\alpha$  increases with $\alpha=1$ 
giving the highest correlation of 0.671 and the lowest MAD value 0.418 in the test data,
indicating minimal deviation between observed and predicted values. 

It's worth noting that the two-stage approach have not produced better predictions for this dataset; 
this is because of the low marker effect variance compared to error variance as evident from extremely low estimated values of 
heritability given in Table \ref{tab:data}.
As per our observations presented in Section \ref{SEC:1vs2}, 
one-stage modeling performs well when dealing with data exhibiting low variance in random SNP effects, 
which is the case in this real dataset. 
Consequently, one-stage model analysis with MDPDE proved to deliver desirable prediction accuracy measures.

\begin{table}[!h]
	\color{black}
\centering
\caption{Prediction accuracy and heritability measures obtained by different one-stage and two-stage procedures applied on Maize dataset (test data results).} 
\begin{tabular}{|l|c|c|c|}
	\hline
	\multicolumn{4}{|c|}{One-stage Procedures} \\ \hline
	Method & $\widehat{\rho}$ & MAD &  $\widehat{H}_p$ \\ \hline\hline
RMLA & 0.404 & 0.658 & 0.006138 \\ 
RMLV & 0.117 & 0.748 & 0.014968 \\ \hline
Rob-RMLA & 0.416 & 0.557 & 0.003128 \\ 
Rob-RMLV & 0.421 & 0.421 & 0.016768 \\ \hline
	\multicolumn{4}{|l|}{Proposed MDPDE based approach} \\ 
$\alpha = 0.1$ & 0.525 & 1.106 & 0.003034 \\ 
$\alpha = 0.3$ & 0.527 & 1.879 & 0.003788 \\ 
$\alpha = 0.5$ & 0.548 & 1.943 & 0.003918 \\ 
$\alpha = 0.7$ & 0.553 & 1.704 & 0.003873 \\ 
$\alpha = 1$ & 0.671 & 0.418 & 0.003074 \\ \hline
\end{tabular}
~~~~~
\begin{tabular}{|l|c|c|c|}
	\hline
	\multicolumn{4}{|c|}{Two-stage Procedures} \\ \hline				
	Method & $\widehat{\rho}$ & MAD &  $\widehat{H}_p$ \\ \hline\hline
Rob1 & 0.622 & 0.377 & 0.00788 \\ 
Rob2 & 0.668 & 0.673 & 0.00762 \\ \hline
	\multicolumn{4}{|l|}{Proposed MDPDE based approach} \\ 
$\alpha = 0.1$ & 0.390 & 1.697 & 0.00576 \\ 
$\alpha = 0.3$ & 0.378 & 1.827 & 0.00580 \\ 
$\alpha = 0.5$ & 0.303 & 1.247 & 0.00584 \\ 
$\alpha = 0.7$ & 0.407 & 1.207 & 0.00593 \\ 
$\alpha = 1$   & 0.410 & 1.087 & 0.00595 \\ \hline
	\multicolumn{4}{c}{} \\
	\multicolumn{4}{c}{} \\
\end{tabular}
\label{tab:data}		
\end{table}

Further, as the heritability estimate in both one-stage and two-stage model analyses tends toward zero,
it also suggests that the variability of this phenotypic trait within the tested maize population is mostly due to environmental factors and not under genetic control.

\section{Conclusions}

In this paper, we have devised robust estimation techniques for genomic prediction based on an underlying linear mixed model. 
In pursuit of more robust estimation, our proposal also employs the popular MDPDE of the parameter of mixed models 
in two distinct modeling structure with one and two-stage approaches. 
We have compared our proposed methods to existing approaches using estimated marker effects in simulation experiments and predicted trait values in a real dataset. 
Remarkably, for both pure and contaminated data, the two-stage modeling structure consistently outperformed other approaches in terms of prediction accuracy
whenever the random SNP-marker effects have greater variance compared to the random noise in the model. 
Our simulation study across various experimental setups demonstrates the sensitivity of results to the parameter $\alpha$ in the MDPDE based proposal
with increasing value of $\alpha$ enhancing the accuracy and robustness of parameter and marker-effect estimation, 
ultimately yielding the more desirable predictive accuracy measure. 
Therefore, our comprehensive analysis establishes that, with a high correlation value and minimal deviation, 
the proposed MDPDE-based approaches with $\alpha=1$ proves to be the optimal choice for both one-stage and two-stage model structures
offering more accurate and reliable gnomic prediction and heritability estimation compared to existing approaches.

We would like to point out that the MDPDE indeed robustifies the maximum likelihood (ML) estimator
and thus our approaches utilize a robust ML-type estimator of variance parameters to make genomic prediction under noisy data.
Although this approach is seen to produce better prediction performances compared to the existing competitors,
the ML-type variance estimators are known to yield biased variance estimates under linear mixed models
while the restricted ML (REML) estimator yields an improved variance estimate. 
Unfortunately, however, when focusing on robustness against data contamination, there is still no method available in the literature 
for producing a robust REML-type variance estimators from possibly contaminated data under the linear mixed models. 
It would be a significantly important future research work to develop such a suitable robust REML-type estimation procedure 
under general linear mixed models  and subsequently incorporate it within our proposed approaches 
to further improve the accuracy of robust genomic prediction.

The promising results and insights derived from this research pave the way for many other exciting future prospects 
in the field of linear mixed models and genotypic-phenotypic effect estimation. 
This could involve exploring and devising new techniques that adapt dynamically to data quality and contamination levels,
e.g., data-driven algorithm for selecting optimum value of the tuning parameter $\alpha$.
The basic heritability measure $\widehat{H}_p$ can also be improved further accounting for selection, 
along the line of, e.g., Schmidt et al.~(2019) \cite{schmidt2019heritability} and Feldmann et al.~(2023) \cite{feldmann2022complex}. 
All these would eventually lead to comprehensive understanding of the intricate relationship between genetic factors and traits.

We also hope to prepare an \texttt{R}-package for the proposed MDPDE-based procedures for genomic prediction and heritability estimation
for dissemination among scientists and practitioners. 
They would then be able to apply our proposals easily in new experimental datasets 
to get more accurate and stable results, even in the presence of any possible noises and/or contamination, 
eventually leading to new insights and innovations. 

\appendix
{\color{black}

\section{Numerical Results obtained on the training datasets}

We have presented all the numerical results obtained under the training data in Tables \ref{tab:1A}--\ref{tab:dataA}.
The relative performances of different one-stage and two-stage methods are again the same as observed in the test data,
although their individual performances are much improved for the training data than those seen under test data.
This is intuitively expected as the models are fitted with the training data. 
However, most importantly, our proposed MDPDE based methods perform quite closely under both their training data as well as the test data.

\begin{table}[!h]
	\color{black}
	\centering
	\caption{Prediction accuracy measures obtained from different methods under the one-stage approach with artificial datasets 
		(training data).}
		\begin{tabular}{|l|c|c|c|c|c|c|}
			\hline
			\multirow{2}{*}{Methods} &  \multicolumn{2}{c|}{Pure Data}  &  \multicolumn{2}{c|}{Random Contamination} & \multicolumn{2}{c|}{Block Contamination} \\
			& $\widehat{\rho}$ & MAD & $\widehat{\rho} $ & MAD & $ \widehat{\rho} $ & MAD\\
			\hline
RMLA & 0.366 & 37.77 & 0.238 & 35.70 & 0.235 & 52.87 \\ 
RMLV & 0.687 & 4.11 & 0.544 & 4.107 & 0.441 & 9.56 \\ \hline
Rob-RMLA & 0.410 & 7.50 & 0.588 & 6.77 & 0.472 & 13.64 \\ 
Rob-RMLV & 0.648 & 5.46 & 0.650 & 4.97 & 0.465 & 27.67 \\ \hline
			\multicolumn{7}{|l|}{Proposed MDPDE based approach} \\ 
$\alpha = 0.1$ & 0.767 & 4.34 & 0.723 & 4.48 & 0.623 & 8.02 \\ 
$\alpha = 0.3$ & 0.745 & 4.37 & 0.766 & 4.36 & 0.666 & 6.38 \\ 
$\alpha = 0.5$ & 0.737 & 5.67 & 0.734 & 4.67 & 0.534 & 6.07 \\ 
$\alpha = 0.7$ & 0.710 & 4.37 & 0.710 & 4.30 & 0.730 & 6.04 \\ 
$\alpha = 1$ & 0.813 & 4.06 & 0.774 & 4.04 & 0.745 & 5.37 \\ \hline
		\end{tabular}
		\label{tab:1A}
\end{table}

\begin{table}[!h]
	\centering
	\color{black}
	\caption{Prediction accuracy measures obtained from different methods under the two-stage approach with artificial datasets 
		(training data).}
		\begin{tabular}{|l|c|c|c|c|c|c|}
			\hline
			\multirow{2}{*}{Methods} &  \multicolumn{2}{c|}{Pure Data}  &  \multicolumn{2}{c|}{Random Contamination} & \multicolumn{2}{c|}{Block Contamination} \\
			& $\widehat{\rho}$ & MAD & $\widehat{\rho} $ & MAD & $ \widehat{\rho} $ & MAD\\
			\hline
Rob1 & 0.756 & 5.43 & 0.788 & 4.07 & 0.772 & 3.64 \\ 
Rob2 & 0.808 & 5.87 & 0.772 & 4.87 & 0.765 & 3.37 \\ \hline
			\multicolumn{7}{|l|}{Proposed MDPDE based approach} \\ 
$\alpha = 0.1$ & 0.837 & 4.34 & 0.803 & 4.04 & 0.799 & 3.22 \\ 
$\alpha = 0.3$ & 0.865 & 4.37 & 0.820 & 3.48 & 0.796 & 3.38 \\ 
$\alpha = 0.5$ & 0.877 & 3.67 & 0.805 & 3.67 & 0.802 & 3.27 \\ 
$\alpha = 0.7$ & 0.840 & 3.77 & 0.830 & 3.37 & 0.814 & 3.33 \\ 
$\alpha = 1$   & 0.884 & 3.07 & 0.833 & 3.30 & 0.823 & 3.18 \\ \hline
		\end{tabular}
		\label{tab:2A}
\end{table}

\clearpage

\begin{table}[!h]
	\centering
	\color{black}{
		\caption{The average value and the MSD of heritability estimators obtained from different one- and two-stage procedures 
			applied on pure and contaminated artificial (training) data.}
		\resizebox{0.95\columnwidth}{!}{
			\begin{tabular}{|l|rr|rr|rr|}
				\hline
				{Methods} & \multicolumn{2}{|c|}{{Pure Data}} & \multicolumn{2}{|c|}{{Random Contamination}} & \multicolumn{2}{|c|}{{Block Contamination}} \\ 
				& {Average} & {MSD} & {Average} & {MSD} & {Average} & {MSD} \\ \hline
				\multicolumn{7}{|c|}{\textbf{One-stage procedures}} \\ \hline  
RMLA & 0.4543 & $6.0368 \times 10^{-4}$ & 0.3181 & $1.4584 \times 10^{-3}$ & 0.3018 & $1.5856 \times 10^{-3}$ \\ 
RMLV & 0.5737 & $1.5951 \times 10^{-4}$ & 0.5492 & $2.274 \times 10^{-4}$ & 0.5175 & $3.3306 \times 10^{-4}$ \\ \hline
Rob-RMLA & 0.6345 & $4.2902 \times 10^{-5}$ & 0.6190 & $6.561 \times 10^{-5}$ & 0.5939 & $1.1257 \times 10^{-4}$ \\ 
Rob-RMLV & 0.6861 & $1.9321 \times 10^{-5}$ & 0.6332 & $4.4622 \times 10^{-5}$ & 0.6303 & $4.858 \times 10^{-5}$ \\ \hline
				\multicolumn{7}{|l|}{Proposed MDPDE based approach} \\
$\alpha = 0.1$ & 0.6927 & $5.329 \times 10^{-7}$ & 0.6570 & $1.849 \times 10^{-5}$ & 0.6373 & $3.9312 \times 10^{-5}$ \\ 
$\alpha = 0.3$ & 0.7515 & $2.6522 \times 10^{-5}$ & 0.7229 & $5.2441 \times 10^{-6}$ & 0.6408 & $3.5046 \times 10^{-5}$ \\ 
$\alpha = 0.5$ & 0.7080 & $6.4 \times 10^{-7}$ & 0.6760 & $5.76 \times 10^{-6}$ & 0.6529 & $2.2184 \times 10^{-5}$ \\ 
$\alpha = 0.7$ & 0.7306 & $9.3636 \times 10^{-6}$ & 0.7104 & $1.0816 \times 10^{-6}$ & 0.6976 & $5.76 \times 10^{-8}$ \\ 
$\alpha = 1$ & 0.6974 & $5.76 \times 10^{-8}$ & 0.6633 & $1.3468 \times 10^{-5}$ & 0.7038 & $1.444 \times 10^{-7}$ \\ \hline
\hline
				\multicolumn{7}{|c|}{\textbf{Two-stage procedures}} \\ \hline  
Rob1 & 0.7796 & $6.3361 \times 10^{-5}$ & 0.7538 & $2.8944 \times 10^{-5}$ & 0.7630 & $3.969 \times 10^{-5}$ \\ 
Rob2 & 0.7906 & $8.2083 \times 10^{-5}$ & 0.7745 & $5.5502 \times 10^{-5}$ & 0.7350 & $1.225 \times 10^{-5}$ \\ \hline
				\multicolumn{7}{|l|}{Proposed MDPDE based approach} \\
$\alpha = 0.1$ & 0.7658 & $4.4329 \times 10^{-5}$ & 0.7492 & $2.4206 \times 10^{-5}$ & 0.6850 & $5.625 \times 10^{-5}$ \\ 
$\alpha = 0.3$ & 0.8100 & $1.21 \times 10^{-4}$ & 0.8003 & $1.006 \times 10^{-4}$ & 0.7280 & $7.84 \times 10^{-6}$ \\ 
$\alpha = 0.5$ & 0.8281 & $1.64 \times 10^{-4}$ & 0.8010 & $1.02 \times 10^{-4}$ & 0.7910 & $8.281 \times 10^{-5}$ \\ 
$\alpha = 0.7$ & 0.8435 & $2.0592 \times 10^{-4}$ & 0.8218 & $1.4835 \times 10^{-4}$ & 0.8030 & $1.0609 \times 10^{-4}$ \\ 
$\alpha = 1$ & 0.8670 & $2.7889 \times 10^{-4}$ & 0.8340 & $1.7956 \times 10^{-4}$ & 0.8220 & $1.4884 \times 10^{-4}$ \\ \hline
\hline
			\end{tabular}
	}		}
	\label{tab:3A1}
\end{table}

\begin{table}[!h]
	\color{black}
	\centering
	\caption{Prediction accuracy and heritability measures obtained by different one-stage and two-stage procedures applied on Maize dataset (training data results).} 
	\begin{tabular}{|l|c|c|c|}
		\hline
		\multicolumn{4}{|c|}{One-stage Procedures} \\ \hline
		Method & $\widehat{\rho}$ & MAD &  $\widehat{H}_p$ \\ \hline\hline
RMLA & 0.561 & 0.591 & 0.003771 \\ 
RMLV & 0.047 & 0.581 & 0.01640 \\  \hline
Rob-RMLA & 0.577 & 0.590 & 0.003561 \\ 
Rob-RMLV & 0.554 & 0.474 & 0.016401 \\ \hline
		\multicolumn{4}{|l|}{Proposed MDPDE based approach} \\ 
$\alpha = 0.1$ & 0.664 & 1.139 & 0.000666 \\ 
$\alpha = 0.3$ & 0.6837 & 1.12 & 0.001422 \\ 
$\alpha = 0.5$ & 0.689 & 1.172 & 0.001351 \\ 
$\alpha = 0.7$ & 0.716 & 1.537 & 0.001306 \\ 
$\alpha = 1$ & 0.82 & 0.453 & 0.000706 \\ \hline
	\end{tabular}
	~~~~~
	\begin{tabular}{|l|c|c|c|}
		\hline
		\multicolumn{4}{|c|}{Two-stage Procedures} \\ \hline				
		Method & $\widehat{\rho}$ & MAD &  $\widehat{H}_p$ \\ \hline\hline
Rob1 & 0.726 & 2.430 & 0.002310 \\ 
Rob2 & 0.726 & 2.420 & 0.002050 \\ \hline
		\multicolumn{4}{|l|}{Proposed MDPDE based approach} \\ 
$\alpha = 0.1$ & 0.294 & 0.750 & 0.000196 \\ 
$\alpha = 0.3$ & 0.282 & 1.880 & 0.000234 \\ 
$\alpha = 0.5$ & 0.307 & 2.300 & 0.000266 \\ 
$\alpha = 0.7$ & 0.311 & 2.260 & 0.000360 \\ 
$\alpha = 1$ & 0.314 & 2.140 & 0.000384 \\ \hline
		\multicolumn{4}{c}{} \\
		\multicolumn{4}{c}{} \\
	\end{tabular}
	\label{tab:dataA}		
\end{table}

}

\bigskip\newpage
\noindent
\textbf{Acknowledgments:}\\
Authors wish to thank the Associate Editor and two anonymous reviewers for their careful reading of  
the manuscript and constructive suggestions to improve the paper. 
This research is supported by a SRG research grant no. SRG/2020/000072 from the Science and Engineering Research Board (SERB), Government of India, India.

\bibliographystyle{acm}
\bibliography{ref}

\begin{thebibliography}{10}

\bibitem{barbosa2005selection}
{\sc Barbosa, M. H.~P., de~Resende, M. D.~V., Bressiani, J.~A., da~Silveira, L.
  C.~I., Peternelli, L.~A., et~al.}
\newblock Selection of sugarcane families and parents by reml/blup.
\newblock {\em Crop Breeding and Applied Technology 5}, 4 (2005), 443.

\bibitem{basu1998robust}
{\sc Basu, A., Harris, I.~R., Hjort, N.~L., and Jones, M.}
\newblock Robust and efficient estimation by minimising a density power
  divergence.
\newblock {\em Biometrika 85}, 3 (1998), 549--559.

\bibitem{basu2011statistical}
{\sc Basu, A., Shioya, H., and Park, C.}
\newblock {\em Statistical inference: the minimum distance approach}.
\newblock CRC press, 2011.

\bibitem{bernal2016outlier}
{\sc Bernal-Vasquez, A.-M., Utz, H.-F., and Piepho, H.-P.}
\newblock Outlier detection methods for generalized lattices: a case study on
  the transition from anova to reml.
\newblock {\em Theoretical and Applied Genetics 129\/} (2016), 787--804.

\bibitem{buhlmann2014high}
{\sc B{\"u}hlmann, P., Kalisch, M., and Meier, L.}
\newblock High-dimensional statistics with a view toward applications in
  biology.
\newblock {\em Annual Review of Statistics and Its Application 1\/} (2014),
  255--278.

\bibitem{chen2009fast}
{\sc Chen, G.~K., Marjoram, P., and Wall, J.~D.}
\newblock Fast and flexible simulation of dna sequence data.
\newblock {\em Genome research 19}, 1 (2009), 136--142.

\bibitem{de2019package}
{\sc de~Mendiburu, F., and de~Mendiburu, M.~F.}
\newblock Package ‘agricolae’.
\newblock {\em R Package, version 1}, 3 (2019), 1143--49.

\bibitem{estaghvirou2014influence}
{\sc Estaghvirou, S. B.~O., Ogutu, J.~O., and Piepho, H.-P.}
\newblock Influence of outliers on accuracy estimation in genomic prediction in
  plant breeding.
\newblock {\em G3: Genes, Genomes, Genetics 4}, 12 (2014), 2317--2328.

\bibitem{estaghvirou2015genetic}
{\sc Estaghvirou, S. B.~O., Ogutu, J.~O., and Piepho, H.-P.}
\newblock How genetic variance and number of genotypes and markers influence
  estimates of genomic prediction accuracy in plant breeding.
\newblock {\em Crop Science 55}, 5 (2015), 1911--1924.

\bibitem{falke2014genome}
{\sc Falke, K.~C., Mahone, G.~S., Bauer, E., Haseneyer, G., Miedaner, T.,
  Breuer, F., and Frisch, M.}
\newblock Genome-wide prediction methods for detecting genetic effects of donor
  chromosome segments in introgression populations.
\newblock {\em BMC genomics 15\/} (2014), 1--10.

\bibitem{feldmann2022complex}
{\sc Feldmann, M.~J., Covarrubias-Pazaran, G., and Piepho, H.-P.}
\newblock Complex traits and candidate genes: estimation of genetic variance
  components across multiple genetic architectures.
\newblock {\em G3: Genes, Genomes, Genetics 13}, 9 (2023), jkad148.

\bibitem{gaynor2021alphasimr}
{\sc Gaynor, R.~C., Gorjanc, G., and Hickey, J.~M.}
\newblock Alphasimr: an r package for breeding program simulations.
\newblock {\em G3 11}, 2 (2021), jkaa017.

\bibitem{ghosh2019robust}
{\sc Ghosh, A.}
\newblock Robust inference under the beta regression model with application to
  health care studies.
\newblock {\em Statistical methods in medical research 28}, 3 (2019), 871--888.

\bibitem{ghosh2013robust}
{\sc Ghosh, A., and Basu, A.}
\newblock Robust estimation for independent non-homogeneous observations using
  density power divergence with applications to linear regression.

\bibitem{ghosh2015robust}
{\sc Ghosh, A., and Basu, A.}
\newblock Robust estimation for non-homogeneous data and the selection of the
  optimal tuning parameter: the density power divergence approach.
\newblock {\em Journal of Applied Statistics 42}, 9 (2015), 2056--2072.

\bibitem{ghosh2016robust}
{\sc Ghosh, A., and Basu, A.}
\newblock Robust estimation in generalized linear models: the density power
  divergence approach.
\newblock {\em Test 25\/} (2016), 269--290.

\bibitem{giraud2021introduction}
{\sc Giraud, C.}
\newblock {\em Introduction to high-dimensional statistics}.
\newblock CRC Press, 2021.

\bibitem{hofheinz2014heteroscedastic}
{\sc Hofheinz, N., and Frisch, M.}
\newblock Heteroscedastic ridge regression approaches for genome-wide
  prediction with a focus on computational efficiency and accurate effect
  estimation.
\newblock {\em G3: Genes, Genomes, Genetics 4}, 3 (2014), 539--546.

\bibitem{koller2013robust}
{\sc Koller, M.}
\newblock {\em Robust estimation of linear mixed models}.
\newblock PhD thesis, ETH Zurich, 2013.

\bibitem{koller2016robustlmm}
{\sc Koller, M.}
\newblock robustlmm: an r package for robust estimation of linear mixed-effects
  models.
\newblock {\em Journal of statistical software 75\/} (2016), 1--24.

\bibitem{lourencco2020robust}
{\sc Louren{\c{c}}o, V.~M., Ogutu, J.~O., and Piepho, H.-P.}
\newblock Robust estimation of heritability and predictive accuracy in plant
  breeding: evaluation using simulation and empirical data.
\newblock {\em BMC genomics 21}, 1 (2020), 1--18.

\bibitem{mohring2009comparison}
{\sc M{\"o}hring, J., and Piepho, H.-P.}
\newblock Comparison of weighting in two-stage analysis of plant breeding
  trials.
\newblock {\em Crop Science 49}, 6 (2009), 1977--1988.

\bibitem{muleta2019optimizing}
{\sc Muleta, K.~T., Pressoir, G., and Morris, G.~P.}
\newblock Optimizing genomic selection for a sorghum breeding program in haiti:
  A simulation study.
\newblock {\em G3: Genes, Genomes, Genetics 9}, 2 (2019), 391--401.

\bibitem{piepho2012efficient}
{\sc Piepho, H., Ogutu, J., Schulz-Streeck, T., Estaghvirou, B., Gordillo, A.,
  and Technow, F.}
\newblock Efficient computation of ridge-regression best linear unbiased
  prediction in genomic selection in plant breeding.
\newblock {\em Crop Science 52}, 3 (2012), 1093--1104.

\bibitem{piepho2009ridge}
{\sc Piepho, H.-P.}
\newblock Ridge regression and extensions for genomewide selection in maize.
\newblock {\em Crop Science 49}, 4 (2009), 1165--1176.

\bibitem{piepho2012stage}
{\sc Piepho, H.-P., M{\"o}hring, J., Schulz-Streeck, T., and Ogutu, J.~O.}
\newblock A stage-wise approach for the analysis of multi-environment trials.
\newblock {\em Biometrical Journal 54}, 6 (2012), 844--860.

\bibitem{resende2016software}
{\sc Resende, M. D. V.~d.}
\newblock Software selegen-reml/blup: a useful tool for plant breeding.
\newblock {\em Crop Breeding and Applied Biotechnology 16\/} (2016), 330--339.

\bibitem{saraceno2020robust}
{\sc Saraceno, G., Ghosh, A., Basu, A., and Agostinelli, C.}
\newblock Robust estimation under linear mixed models: The minimum density
  power divergence approach.
\newblock {\em arXiv preprint arXiv:2010.05593\/} (2020).

\bibitem{schmidt2019heritability}
{\sc Schmidt, P., Hartung, J., Bennewitz, J., and Piepho, H.-P.}
\newblock Heritability in plant breeding on a genotype-difference basis.
\newblock {\em Genetics 212}, 4 (2019), 991--1008.

\bibitem{shen2013novel}
{\sc Shen, X., Alam, M., Fikse, F., and R{\"o}nneg{\aa}rd, L.}
\newblock A novel generalized ridge regression method for quantitative
  genetics.
\newblock {\em Genetics 193}, 4 (2013), 1255--1268.

\bibitem{tanaka2020simple}
{\sc Tanaka, E.}
\newblock Simple outlier detection for a multi-environmental field trial.
\newblock {\em Biometrics 76}, 4 (2020), 1374--1382.

\bibitem{uffelmann2021genome}
{\sc Uffelmann, E., Huang, Q.~Q., Munung, N.~S., De~Vries, J., Okada, Y.,
  Martin, A.~R., Martin, H.~C., Lappalainen, T., and Posthuma, D.}
\newblock Genome-wide association studies.
\newblock {\em Nature Reviews Methods Primers 1}, 1 (2021), 59.

\bibitem{wainwright2019high}
{\sc Wainwright, M.~J.}
\newblock {\em High-dimensional statistics: A non-asymptotic viewpoint},
  vol.~48.
\newblock Cambridge university press, 2019.

\bibitem{westhues2021prediction}
{\sc Westhues, C.~C., Mahone, G.~S., da~Silva, S., Thorwarth, P., Schmidt, M.,
  Richter, J.-C., Simianer, H., and Beissinger, T.~M.}
\newblock Prediction of maize phenotypic traits with genomic and environmental
  predictors using gradient boosting frameworks.
\newblock {\em Frontiers in plant science 12\/} (2021), 699589.

\bibitem{zhu2020statistical}
{\sc Zhu, H., and Zhou, X.}
\newblock Statistical methods for snp heritability estimation and partition: A
  review.
\newblock {\em Computational and Structural Biotechnology Journal 18\/} (2020),
  1557--1568.

\end{thebibliography}

\end{document}